\documentclass[preprint,aps,superscriptaddress,showpacs]{revtex4}
\pdfoutput=1
\usepackage{amsmath}
\usepackage{mathtools}
\usepackage{amsfonts}
\usepackage{grffile}
\usepackage{graphicx}
\usepackage{caption}
\usepackage{subcaption}
\usepackage{dcolumn}
\usepackage{color}
\usepackage{hyperref}
\usepackage[all]{hypcap}
\usepackage{bm}
\usepackage{color}
\usepackage{float}
\usepackage{xspace}
\usepackage{libertine}
\usepackage[makeroom]{cancel}

\newcommand{\bra}[1]{\langle #1|}
\newcommand{\ket}[1]{|#1\rangle}
\newcommand{\braket}[2]{\langle #1|#2\rangle}
\newcommand{\ketbra}[2]{| #1 \rangle \langle #2 |}
\newcommand{\Tr}[1]{\operatorname{Tr}\left[ #1 \right]}

\newcommand{\abs}[1]{\left\vert#1\right\vert}
\newcommand{\trace}[1]{\mbox{Tr}\left( #1 \right)}

\begin{document}

\title{Efficient Quantum Tomography of  Two-Mode Wigner Functions}
\author{Ludmila A. S. Botelho}
\email{lasb@ufmg.br}
\affiliation{Departamento de F\'{i}sica - ICEx - Universidade Federal de Minas Gerais, \\ Av. Pres. Ant\^{o}nio Carlos 6627 - Belo Horizonte, MG, Brazil - 31270-901.}
\author{Reinaldo O. Vianna}
\affiliation{Departamento de F\'{i}sica - ICEx - Universidade Federal de Minas Gerais, \\ Av. Pres. Ant\^{o}nio Carlos 6627 - Belo Horizonte, MG, Brazil - 31270-901.}

\date{\today}

\begin{abstract}

We introduce an efficient method to reconstruct the Wigner function of many-mode continuous variable systems. It is based on convex optimization with semidefinite
programs, and also includes a version of the maximum entropy principle, in order to yield unbiased states. A key ingredient of the proposed approach is the representation
of the state in a truncated Fock basis. As a bonus, the discrete finite representation allows to easily quantify the entanglement.

\end{abstract}



\maketitle

\section{Introduction}

The Wigner function \cite{Wigner1932} is a quasi-probability phase space distribution of wide application in diverse areas as quantum physics, quantum electronics,
quantum chemistry, and signal processing, to mention a few \cite{Review2018}.  In the effervescent field of Quantum Information, it is a tool of paramount importance in the
investigation of  information processing  and quantum entanglement  based on continuous variables (CV) \cite{Review2005}.  Knowing it is tantamount to the
knowledge of the  quantum state, which by its turn carries all the information one can know about the quantum   system. If one can experimentally reconstruct the 
Wigner function, by means of Quantum Tomography \cite{Review2009}, then the probability distribution associated to the measurement of any property can be predicted.

The first approach introduced to perform the Wigner Function Quantum Tomography (WQT) was based on the inverse Radon Transform \cite{Bertrand1987}.
 However, besides having a
high computational cost, it can yield non-physical functions, equivalent to non-positive density operators, due to imperfect measurements. This last drawback can be corrected
by means of Maximum Likelihood approaches (MaxLik) \cite{MaxLik}. Notwithstanding, MaxLik can still be impractical due to the high cost of non-linear optimizations. 
Another difficulty with WQT is that CV quantum states have infinite dimensional Hilbert spaces, therefore one has  a quantum tomography scenario that  in practice is
always informationally incomplete. In this case, the state can not be uniquely determined, {\em i.e.}, there is a family of states compatible
with the available data. In order to single out a state from this family, one can invoke the Maximum Entropy Principle (MaxEnt) \cite{Jaynes}, which establishes that the 
least unbiased state is the one that maximizes the information entropy. Combining MaxLik and MaxEnt  \cite{Buzek} is a strategy superior to a direct inverse Radon Transform, but
is still very computationally demanding. For finite Hilbert spaces, we  previously introduced an efficient tomographic approach based on semidefinite programs (SDP), which can handle
noisy and informationally incomplete measurements \cite{Maciel2009, Maciel2011,Maciel2012} in the spirit of the MaxEnt principle \cite{Douglas}. In this paper, we extend our technique to the case of CV states,  and show that two-mode states can be reconstructed efficiently with very few measurements and post-processing in low cost desktop computers  (say 4 GB and Intel i5 core processor).

The paper is organized as follows. After a quick  review  of the quantum mechanical description of continuous variable systems, in Sec. II, we introduce our quantum tomography technique
in Sec. III.  Our approach is illustrated in Sec. IV, where we reconstruct some typical two-mode CV states, and then we conclude in Sec. V.

\section{Description of Continuous Variable Systems}

\subsection{The Wigner Function}

If $x$ is a random variable, according to classical probability theory  its characteristic function $\Phi_x(t)$ is the expectation value of $\exp{(-itx)}$, with $t$ being a real number, namely,
\begin{equation} \label{Eq-rov-1}
\Phi_x(t)=\langle  \exp{(-itx)} \rangle.
\end{equation}
Out of the characteristic function, we can build the probability density function for $x$ as:
\begin{equation}\label{Eq-rov-2}
F_t(x)=\frac{1}{2\pi}\int_{-\infty}^{+\infty}  {e^{{itx}}\Phi_x(t)dt}.
\end{equation}
 Now we want to build a probability density function for the position ($\hat{q}$) and momentum 
($\hat{p}$) operators, which satisfy the usual commutation relation:
\begin{equation}\label{Eq-rov-3}
[\hat{q},\hat{p}]=i\hbar \mathbb{I}.
\end{equation} 
The characteristic function is formed by the expectation value of an exponential of $\hat{q}$ and $\hat{p}$ analogous to Eq.\ref{Eq-rov-1}, namely:
\begin{equation}\label{Eq-rov-4}
\Phi_{q,p}(u,v)=\trace{\rho e^{-i(u\hat{q}+v\hat{p})/\hbar}},
\end{equation}
where $u/\hbar$ and $v/\hbar$ are real numbers, analogous to $t$, $\rho$ is the quantum state's density matrix, and we used the Born rule to calculate the expectation value. 
The exponential in Eq.\ref{Eq-rov-4} is known as the Weyl operator. The probability density function  associated to $\Phi_{q,p}(u,v)$, analogous to Eq.\ref{Eq-rov-2}, is the Wigner function \cite{Ballentine, Zyc}:
\begin{equation}\label{Eq-rov-5}
W(q,p)=\left(\frac{1}{2\pi\hbar}\right)^2      \int_{-\infty}^{+\infty}\int_{-\infty}^{+\infty} e^{i(uq+vp)/\hbar} \Phi_{q,p}(u,v) du dv.  
\end{equation}
Notice that $q$ and $p$ are the real values assumed by the position and momentum operators and define the quantum phase space.  
Using the identity \cite{Zyc}:
\begin{equation}\label{Eq-rov-6}
e^{-i(u\hat{q}+v\hat{p})/\hbar}=\int_{-\infty}^{+\infty} e^{-iuq^\prime/\hbar}\ket{q+\frac{v}{2}}\bra{q^\prime-\frac{v}{2}}dq^\prime,
\end{equation}
and after some algebra, we can rewrite the Wigner function in its well known form,
\begin{equation}\label{Eq-rov-7}
W(q,p)=\frac{1}{2\pi\hbar} \int_{-\infty}^{+\infty}  \bra{q-\frac{v}{2}}\rho\ket{q+\frac{v}{2}}e^{ivp/\hbar}dv.
\end{equation}Though $W(q,p)$ yields the correct
quantum probabilities,
\begin{equation} \label{Eq-rov-8}
\int_{-\infty}^{+\infty} W(q,p)dp=\bra{q}\rho\ket{q},
\end{equation}
\begin{equation} \label{Eq-rov-9}
\int_{-\infty}^{+\infty} W(q,p)dq=\bra{p}\rho\ket{p},
\end{equation}
it is not positive semidefinite, {\em i.e.}, it is a {\em quasi-probability} density function. Expectation values of observables ($\hat{O}$) can be calculated directly from
the Wigner function as:
\begin{equation}\label{Eq-rov-10}
\trace{\rho \hat{O}}=\int_{-\infty}^{+\infty}\int_{-\infty}^{+\infty}W(q,p) O_W(q,p)dqdp,
\end{equation}
with the Wigner representation of the observable defined as:
\begin{equation}\label{Eq-rov-11}
O_W(q,p)= \int_{-\infty}^{+\infty} \bra{q-\frac{v}{2}}\hat{O}\ket{q+\frac{v}{2}}e^{ivp/\hbar}dv.
\end{equation}
In Classical Physics, expectation values are calculated by means of averages in phase space, and Eq.\ref{Eq-rov-10} allows us to do an analogous calculation in
Quantum Physics.  

It is straightforward to re-derive the Wigner function for a system of $N$ particles, and it reads:
\begin{eqnarray}\label{Eq-rov-12}
W(q_1,p_1,q_2,p_2,\ldots,q_N,p_N)=\left(\frac{1}{2\pi\hbar}\right)^N    \int_{-\infty}^{+\infty}\ldots\int_{-\infty}^{+\infty} e^{i(v_1p_1+v_2p_2+\ldots+v_Np_N)/\hbar}  \\ \nonumber
\bra{q_1-\frac{v_1}{2},q_2-\frac{v_2}{2},\ldots,q_N-\frac{v_N}{2}}\rho\ket{q_1+\frac{v_1}{2},q_2+\frac{v_2}{2},\ldots,q_N+\frac{v_N}{2}}dv_1dv_2\ldots dv_N.
\end{eqnarray}

\subsection{The Inverse Radon Transform}

Consider the canonical transformation defining the quadratures $\hat{q}_\theta$ and $\hat{p}_\theta$:
\begin{eqnarray}\label{Eq-rov-13}
\hat{q}_\theta=\hat{q}\cos{\theta}+\hat{p}\sin{\theta}, \\ \nonumber
\hat{p}_\theta=-\hat{q}\sin{\theta}+\hat{p}\cos{\theta}, \\ \nonumber
[\hat{q}_\theta,\hat{p}_\theta]=[\hat{q},\hat{p}]=i\hbar.
\end{eqnarray}
The probability distribution associated to these quadratures, say
\begin{equation} \label{Eq-rov-14}
pr(q,\theta)=\bra{q_\theta}\rho\ket{q_\theta},
\end{equation}
 can be experimentally determined by means of Homodyne detection \cite{Ballentine, Collett}, and it corresponds to the Radon transform of the Wigner function,
 namely:
 \begin{equation}\label{Eq-rov-15}
\bra{q_\theta}\rho\ket{q_\theta}\! = \frac{1}{2\pi\hbar}\!\! \int  \!\!W(q_{\theta}\!\cos{\theta} - p_{\theta}\!\sin{\theta},q_\theta\! \sin{\theta} + p_{\theta}\!\cos{\theta}) \mathrm{d}p_{\theta}.
\end{equation}
The Wigner function can then be reconstructed by means of the inverse Radon transform \cite{Bertrand1987},
\begin{equation}\label{Eq-rov-16}
W(q,p) = \frac{1}{2 \pi^{2}} \int_{-\infty}^{+\infty} \! \! \!\int_{0}^{\pi} \!\!pr(x,\theta) K(q\cos{\theta} + p\sin{}\theta -x) \mathrm{d}x \mathrm{d}\theta,
\end{equation}
where the integration kernel $K(x)$ is  defined as:
\begin{equation}\label{Eq-rov-17}
K(x)= \frac{1}{2} \int_{-\infty}^{+\infty} \abs{\xi} \exp{(i\xi x)} \mathrm{d}\xi.
\end{equation}   
In the case of a single-mode state, the reconstruction is achieved by  replacing  the kernel $K(x)$ with a regularized numerical approximation,
 obtained by choosing an appropriate cutoff frequency for the integral in Eq.\ref{Eq-rov-17}. It results in a well known algorithm for image reconstruction,
 which is not viable for multi-mode states \cite{Herman1980}, due to technical problems
 like the very large multi-dimensional grid needed for the inversion,  and the difficulty of adjusting the frequency cutoffs for the many modes.

\subsection{Computational Basis and Informationally Complete Measurements}

The quantum harmonic oscillator sets the ground for the choice of both  the computational basis and Information
Complete measurements needed in continuous variable  quantum tomography. To simplify notation, from now on we 
adopt dimensionless operators $\hat{q}$ and $\hat{p}$, then $\hbar=1$ and $[\hat{q},\hat{p}]= i\mathbb{I}$. We start from the annihilation ($\hat{a}$) and creation
($\hat{a}^\dagger$) operators, which satisfy the bosonic commutation relation:
\begin{equation}\label{Eq-rov-20}
[\hat{a},\hat{a}^\dagger]=\mathbb{I}.
\end{equation}
The dimensionless position and momentum operators, or the quadrature operators, are related to them according to:
\begin{eqnarray}\label{Eq-rov-21}
\hat{q}=\frac{\hat{a}^\dagger +\hat{a}}{\sqrt{2}},\\
\hat{p}=\frac{i(\hat{a}^\dagger -\hat{a})}{\sqrt{2}}.
\end{eqnarray}
The dimensionless harmonic oscillator Hamiltonian reads (with frequency $\omega$ and mass $M$ set to 1):
\begin{equation}\label{Eq-rov-22}
\hat{H}=\hat{a}^\dagger \hat{a} +\frac{1}{2}, 
\end{equation}
and its eigenstates are the number states or Fock states, namely,
\begin{equation}\label{Eq-rov-23}
\hat{a}^\dagger \hat{a} \ket{n}=n\ket{n}.
\end{equation}
In coordinate representation, a Fock state reads:
\begin{equation}\label{Eq-rov-23b}
\braket{q}{n}=\psi_n(q)=\left( \frac{\alpha}{\pi^{1/2}2^n n!}\right)^{1/2}\mathcal{H}_n(\alpha q)e^{-\frac{1}{2}\alpha^2 q^2},
\end{equation}
with $\alpha=(M\omega/\hbar)$, and $\mathcal{H}_n$ being the Hermite polynomial of degree $n$.

In contrast to the continuous position and momentum representations, 
\begin{equation} \label{Eq-rov-24}
\mathbb{I}=\int_{-\infty}^{+\infty} \ketbra{q}{q}dq =\int_{-\infty}^{+\infty} \ketbra{p}{p}dp,
\end{equation}
now we have a discrete representation for the CV systems:
\begin{equation} \label{Eq-rov-25}
\mathbb{I}=\sum_{n=0}^{\infty} \ketbra{n}{n}.
\end{equation}
 These three representations are equivalent, and the discrete one is the natural choice for our  Computational Basis.

 We come then to the last ingredient for our quantum tomography, which is the Informationally Complete basis. 
 With the inverse Radon transform (Eq.\ref{Eq-rov-5}), we learned that the quadratures  (Eq.\ref{Eq-rov-13}) form such a basis, namely:
 \begin{equation} \label{Eq-rov-26}
 \mathbb{I}=\frac{2}{\pi}\int_{-\infty}^{+\infty} \int_0^{\pi} \ketbra{q_\theta}{q_\theta} dqd\theta.
 \end{equation}
  Another Informationally Complete basis is formed by the eigenstates of the non-Hermitian annihilation operator, namely,  the coherent states \cite{Zyc}:
 \begin{equation} \label{Eq-rov-27}
 \hat{a}\ket{z}=z\ket{z},  \,\, z=\frac{1}{\sqrt{2}}(q+ip); 
 \end{equation}
 \begin{equation}\label{Eq-rov-28}
  \ket{q,p}\equiv\ket{z}=e^{z\hat{a}^\dagger-z^* \hat{a}}\ket{0}=e^{-\abs{z}^2/2}=\sum_{n=0}^\infty \frac{z^n}{\sqrt{n!}}\ket{n};
  \end{equation}
 \begin{equation}\label{Eq-rov-29}
 \mathbb{I}=\frac{1}{2\pi}\int_{-\infty}^{+\infty} \int_{-\infty}^{+\infty} \ket{q,p}\bra{q,p} dqdp.
  \end{equation}
 The probability distributions associated to the informationally complete bases can be experimentally measured by Homodyne detection \cite{Collett}, in the
 case of quadratures (Eq.\ref{Eq-rov-26}), and by Heterodyne detection, in the case of coherent states (Eq.\ref{Eq-rov-29}) \cite{Collett,Chabaud}.

\section{Quantum Tomography with Informationally Incomplete Data}

Now we take a complete different perspective to the problem of reconstructing the Wigner function.  We focus on the quantum state and consider
its reconstruction as a matrix completion problem \cite{Fazel} .   If some entries of a positive semidefinite matrix is known, it is possible 
to obtain a low rank approximation to it by means of convex optimization \cite{Maciel2009,Candes}.  We first used such idea to estimate entanglement
of unknown mixed states \cite{Maciel2009},  and then further developed it to efficient quantum tomography techniques
based on semidefinite programs \cite{Maciel2011,Maciel2012,Douglas}. Once the quantum state is obtained, the Wigner function can be calculated by
means of Eq.\ref{Eq-rov-12}.

Though we are dealing with CV states and infinite dimensional Hilbert spaces, notice that the computations are always performed with a finite dimensional
representation. In the case of the inverse Random transform, for instance,  we  choose binnings for the quadratures (Eq.\ref{Eq-rov-13}, Eq.\ref{Eq-rov-14}) and cutoffs
for the convolutions (Eq.\ref{Eq-rov-16}, Eq.\ref{Eq-rov-17}). Therefore, for the sake of this discussion, we assume a finite dimensional Fock state representation (Eq.\ref{Eq-rov-25}).

Consider a continuous set of informationally complete  ($\mathcal{IC}$) observables ($\{E(\alpha)\}$) forming a positive operator valued measure (POVM), {\em i.e.},
 it expands the density matrix:
\begin{equation}\label{Eq-rov-18}
\int  E(\alpha) d\alpha=\mathbb{I}, \,\, E(\alpha)\succeq 0.
\end{equation} 
Suppose a few of these observables are actually measured, forming a finite discrete subset  ($\mathcal{I}=\{\langle E_i \rangle = f_i\}\subset\mathcal{IC}$), which corresponds
to the information we have ($\mathcal{I}$) about the state. The following SDP  reconstructs a unbiased  approximation to the 
quantum state  \cite{Douglas}: 

\begin{equation}\label{Eq-rov-19}
\begin{aligned}
\min_{\rho,\Delta_i,\delta} & \sum_{i\in\mathcal{I} }\Delta_i + \delta & \\ 
\text{subject to:} &    & \\
                           &\vert \trace{E_i \rho} - f_i\vert \leq \Delta_i f_i, & \forall i\in \mathcal{I}\\
                           & \trace{ (\mathbb{I}-\sum_{i\in \mathcal{I}} E_i)\rho}\leq \delta, &     \\
                           & \Delta_i\geq 0 , & \forall i\in\mathcal{I} \\
                           & \delta \geq 0 & \\
                           & \trace{\rho}=1, & \\
                           & \rho \succeq 0.& 
\end{aligned}
\end{equation}
Minimization  of the variational parameters $\{\Delta_i\}$ takes account of  the measurement errors associated to the observed frequencies $\{f_i\}$. Minimizing
 the variational parameter
$\delta$ forces the probabilities of the non-measured observables ($\langle E_i \rangle \notin \mathcal{I}$)  to 
be the most uniform as possible, as would be required by the Maximum Entropy Principle \cite{Jaynes}. The second program line adjusts the estimated state
$\rho$ to the observed frequencies. The third program line adjusts the non-measured observables to the MaxEnt principle. Then follows the constraints on
the SDP variables, being  $\rho$ a positive semidefinite trace one operator, and  $\Delta_i$ and $\delta$ are  non-negative real numbers.

\section{Applications}

In this section, we illustrate the application of our tomographic approach (Eq.\ref{Eq-rov-19}) to typical CV states, chosen based on experimental interest and realizability in the
laboratory. The tomographies are performed out of quadratures (Eq.\ref{Eq-rov-26}) or coherent states (Eq.\ref{Eq-rov-29}). The phase space is uniformly sampled, with
$q$ (Eq.\ref{Eq-rov-13}) varying from 5 to -5, and $\theta$ from 0 to $\pi$, in the first case, and $z$ (Eq.\ref{Eq-rov-28}) taken as a  real number varying from 0 to 2, in the second case. The computational basis has dimension 11, which  corresponds to a finite Fock state representation (Eq.\ref{Eq-rov-25}) truncated at   $n=10$ . This choice for $n$ guarantees a numerical precision
of $10^{-8}$ in the expectation values. The calculations were performed using MATLAB (\cite{matlab}), and MOSEK (\cite{mosek}) interfaced with YALMIP (\cite{yalmip}) to solve the SDPs.

\subsection{1-Photon Fock State: Inverse Radon Transform versus SDP}
To illustrate the superior efficiency of SDP tomography over  the commonly used Inverse Radon Transform (IRT), we reconstructed the one-photon Fock state ($\ket{1}$) 
(Eq.\ref{Eq-rov-23b}), in absence of noise.
The quadrature operators were uniformly sampled in  phase space, taking 7 different values for $q$,  and angles spaced  by  $\pi/4$ increments, resulting in 35 projectors. This low amount of operators resulted in a perfect reconstruction via SDP (Fig.1a) . 
\begin{figure}\label{fig:fig1}
	\begin{subfigure}{0.4\textwidth}
		\includegraphics[width=\textwidth]{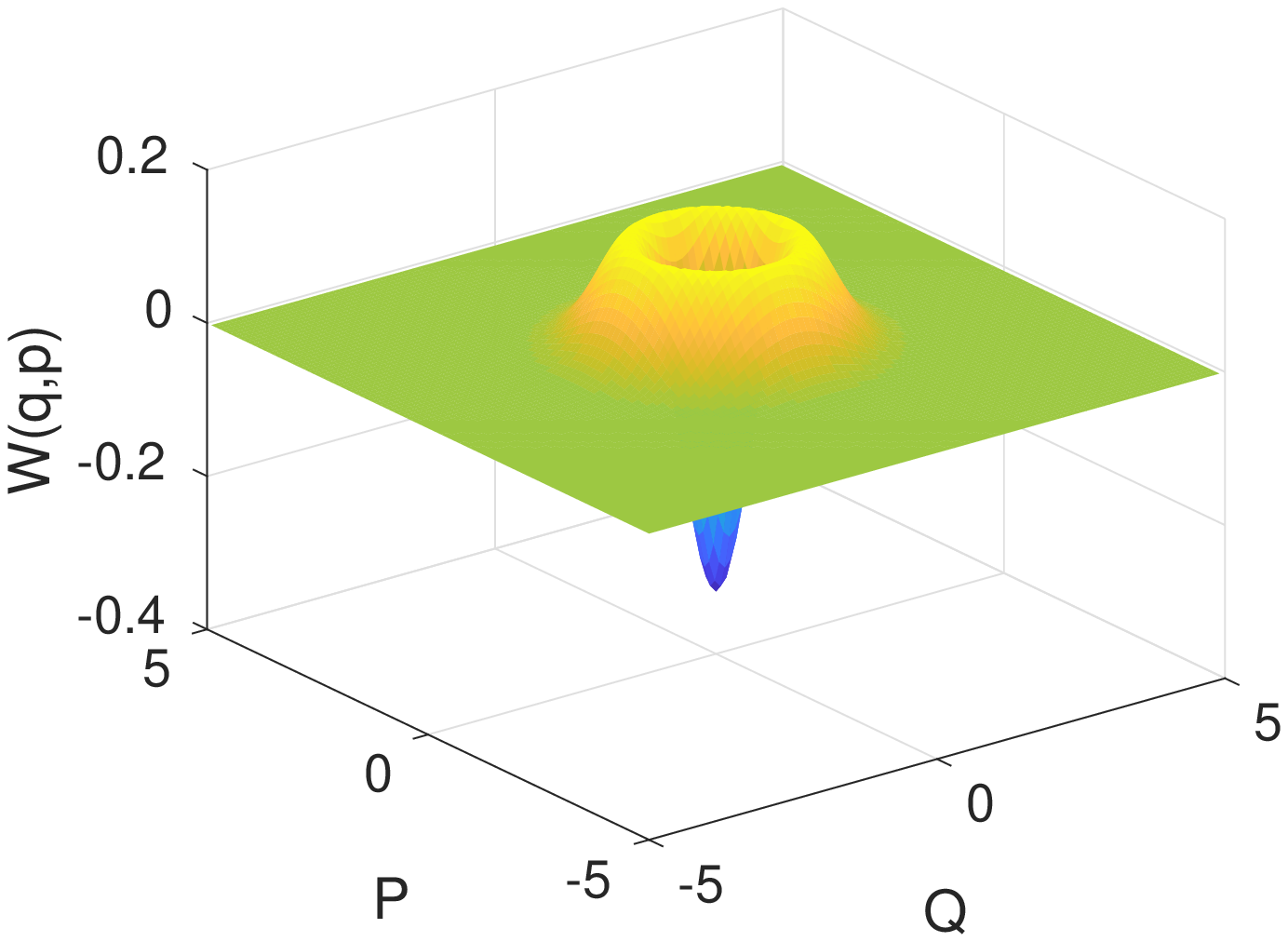}
		\caption{\label{fig:fig1a}}
	\end{subfigure}
	\begin{subfigure}{0.4\textwidth}
		\includegraphics[width=\textwidth]{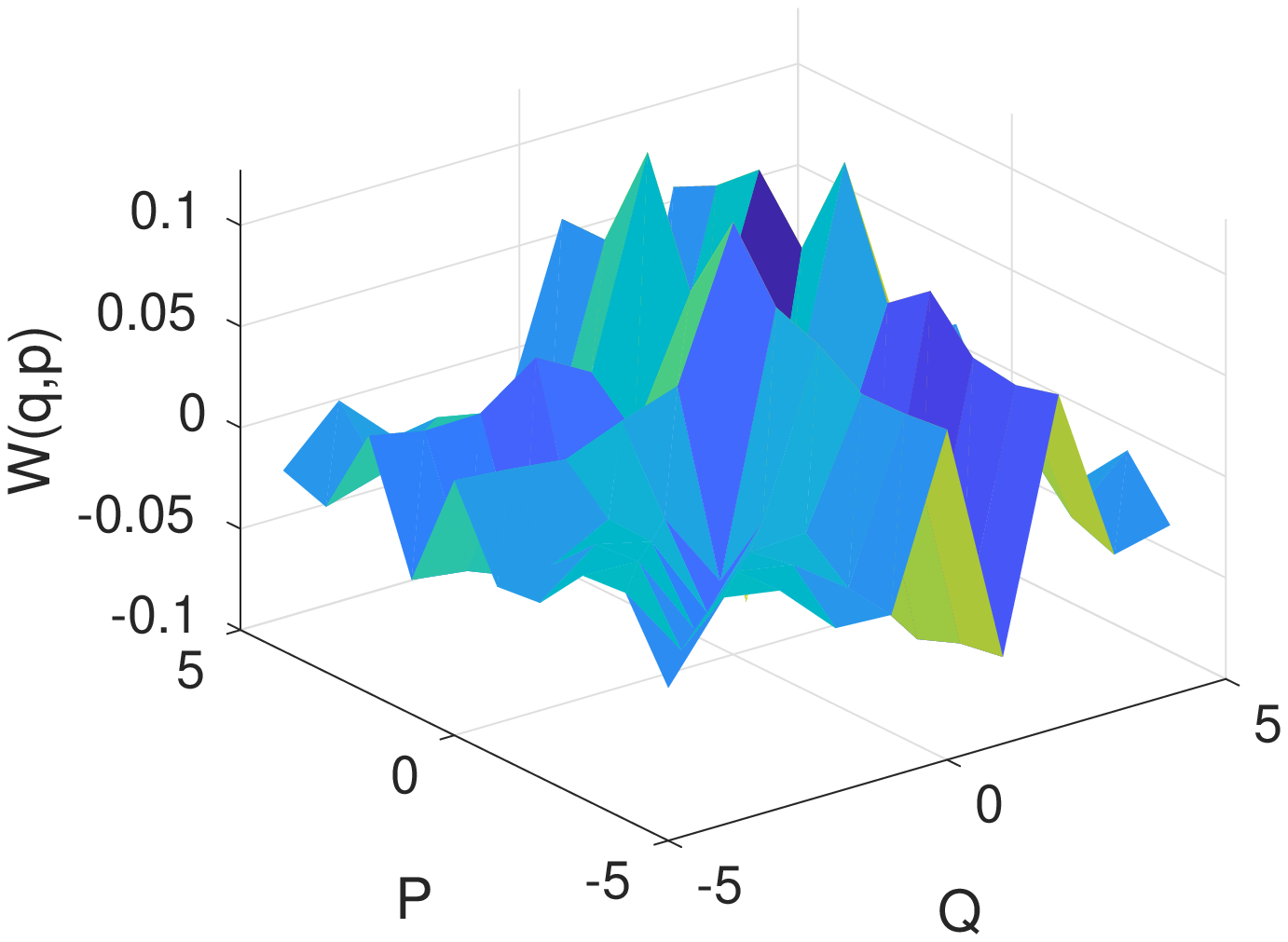}
		\caption{\label{fig:fig1b}}
	\end{subfigure}
	\caption{Wigner Function of the Fock state $\ket{1}$: (a) SDP reconstruction, and  (b) Inverse Radon Transform reconstruction with the same amount of measurements.}
\end{figure}
The resulting Wigner Function obtained via IRT is shown in Fig.\ref{fig:fig1b}. 
As we can see,  with this number of measurements the IRT delivers a very poor result, leading to a non-physical state.
 In Fig.2, we can see the density matrix  reconstructed with the two methods. 
 \begin{figure}
	\begin{subfigure}{0.4\textwidth}
		\includegraphics[width=\textwidth]{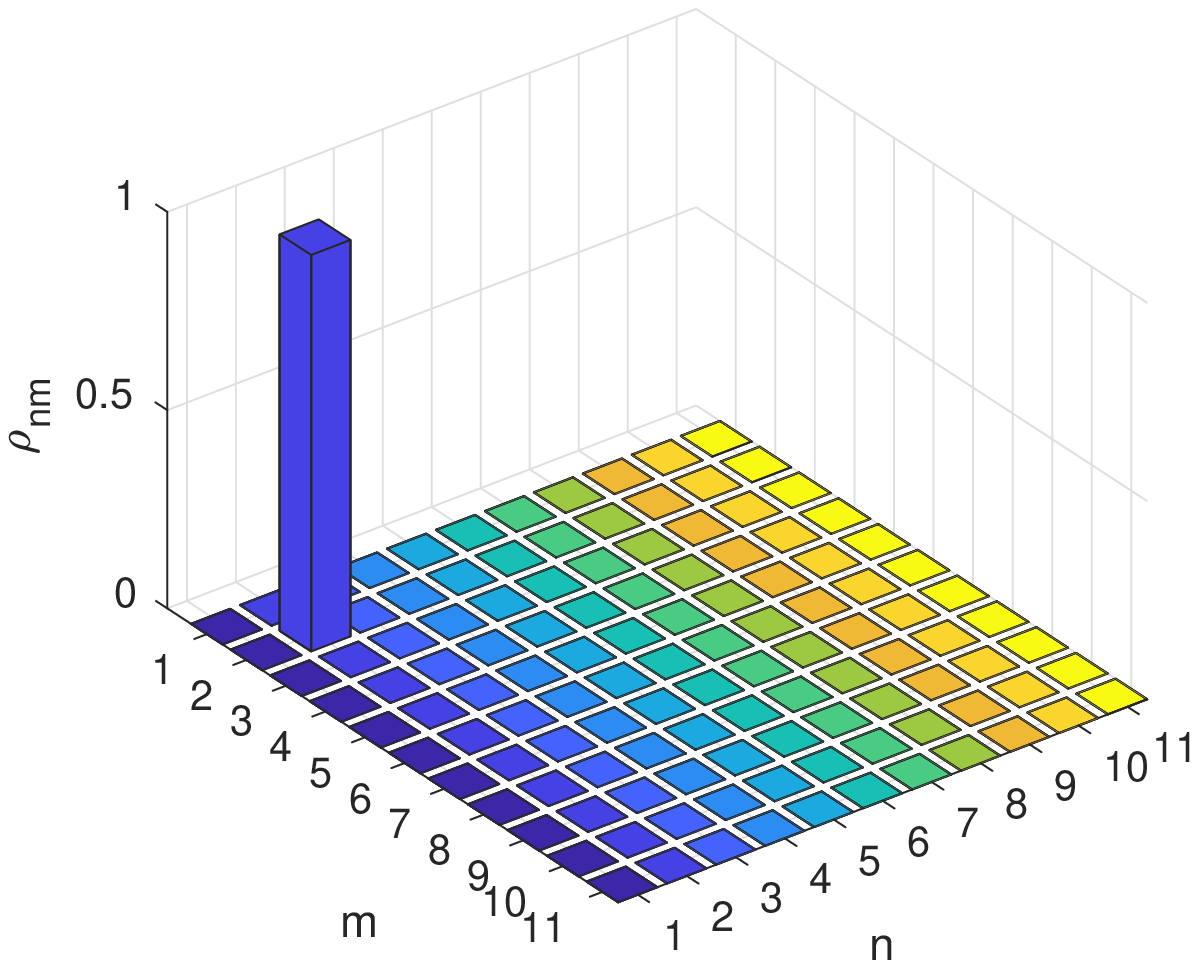}
		\caption{}
		\label{fig:fig2a}
	\end{subfigure}
	\begin{subfigure}{0.4\textwidth}
	\includegraphics[width=\textwidth]{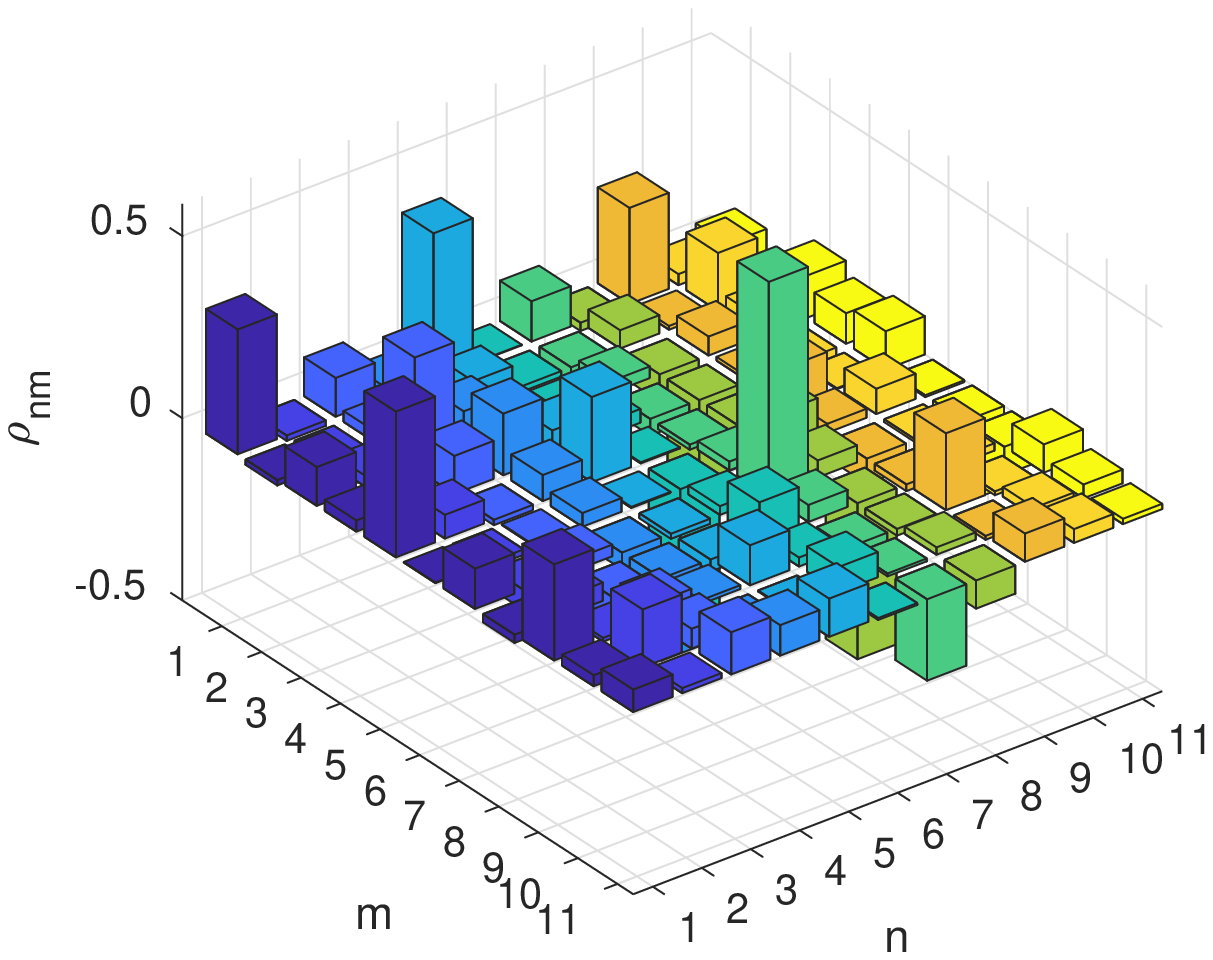}
		\caption{}
		\label{fig:fig2b}
	\end{subfigure}
	\caption{The density matrix  of Fock state $\ket{1}$ reconstructed via (a)  SDP,  and (b) Inverse Radon Transform (IRT). Notice that the IRT reconstruction resulted in negative values in the diagonal.}
\end{figure}

\subsection{Two Mode Entangled States}
Now we reconstruct four different CV two-mode states, relevant as entanglement resources  \cite{Walborn2011}  and also useful in quantum metrology \cite{Mitchell}:
the NOON state in  the  Fock basis, 
 \begin{equation}
 \ket{\Psi}_{NOON}=\frac{1}{\sqrt{2}}(\ket{1}_1\ket{0}_2+\ket{0}_1\ket{1}_2),
 \end{equation}
formed by a one-photon mode and a vacuum mode;
the  \emph{Hermite-Gauss state}:
\begin{equation}
\Phi(q_1,q_2)=\frac{A_n}{\sqrt{\sigma_+ \sigma_-}}\mathcal{H}_n \left( \frac{q_1+q_2}{\sqrt{2}\sigma_+}\right)e^{\frac{-(q_1+q_2)^2}{4\sigma_+^2}}e^{\frac{-(q_1-q_2)^2}{4\sigma_-^2}}
\end{equation}
where $\mathcal{H}_n$ is the Hermite polynomial of degree $n$, $A_n$ is a normalization factor and we set  $n=1$, $\sigma_+ =1$ and $\sigma_-=0.5$;
 the \emph{two-mode squeezed vacuum state}, which in position representation reads
\begin{equation}
 \Psi(q_1,q_2)=\frac{1}{\sqrt{\pi}}\operatorname{exp}\left[-\frac{1}{4}e^{2\zeta}(q_1+q_2)^2-\frac{1}{4}e^{-2\zeta}(q_1-q_2)^2\right],
\end{equation}
while represented  in the Fock basis as
\begin{equation}
\ket{\Psi}=\sqrt{1-\lambda^2}\sum_{n=0}^{\infty}\lambda^{n} \ket{n}\ket{n},
\end{equation}
with  $\lambda=\tanh{\zeta}$ and $\zeta$ is the squeezed parameter, set to $\zeta=0.2$;
and finally a  mixed state in the coherent states basis, the \emph{Dephased Cat},
 \begin{equation}
 \rho=N(\alpha,p)\{\ketbra{\alpha,\alpha}{\alpha,\alpha}+\ketbra{-\alpha,-\alpha}{-\alpha,-\alpha}-(1-p)(\ketbra{\alpha,\alpha}{-\alpha,-\alpha}+\ketbra{-\alpha,-\alpha}{\alpha,\alpha})\}
 \end{equation}
with  $N(\alpha,p)$  a normalization constant, and we set  $\alpha = 1$ and $p=0.5$.
\begin{figure}
	\begin{subfigure}{0.4\textwidth}\
		\includegraphics[width=\textwidth]{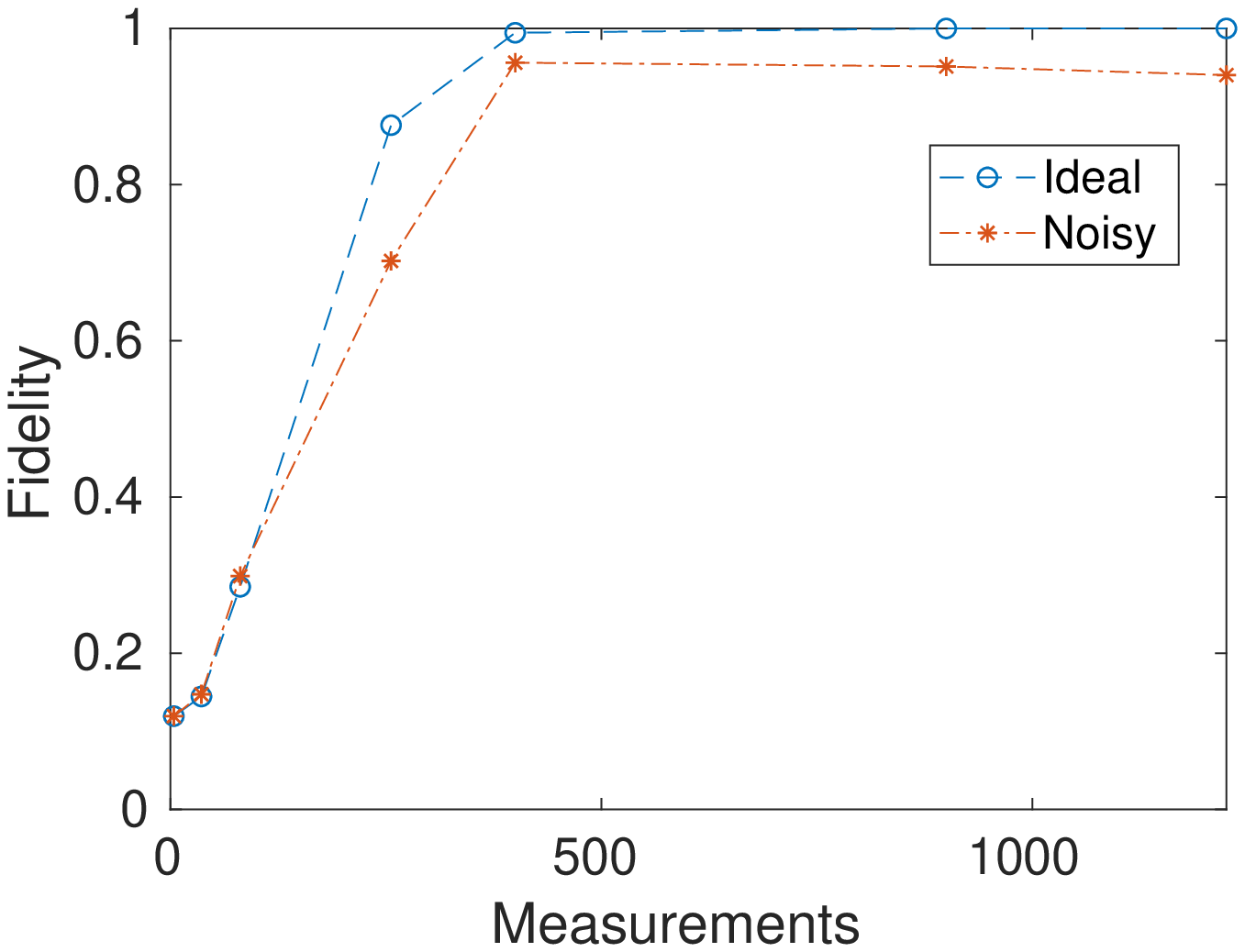}
		\caption{}
		\label{fig:fig4a}
	\end{subfigure}
	\begin{subfigure}{0.4\textwidth}
		\includegraphics[width=\textwidth]{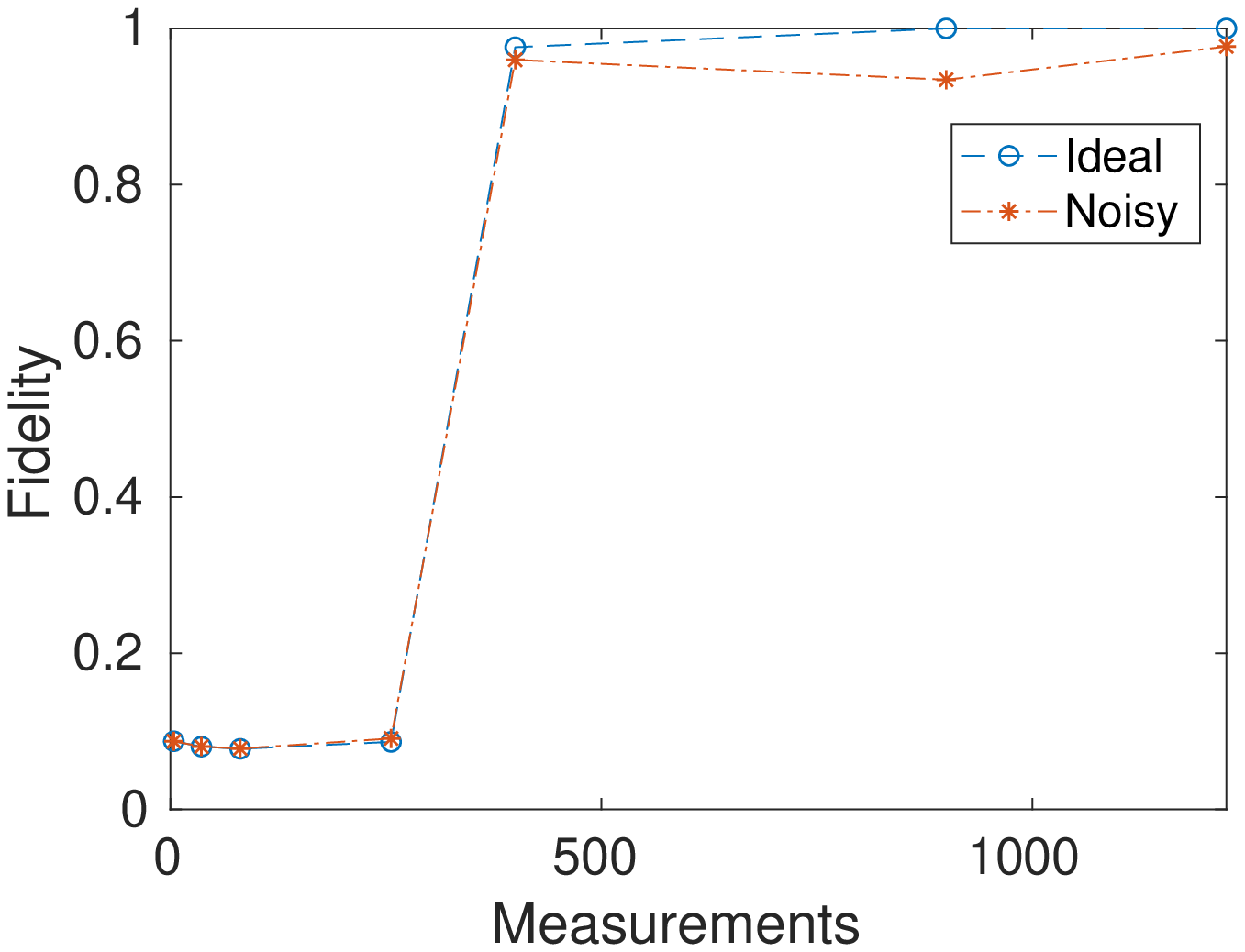}
		\caption{}
		\label{fig:fig4b}
	\end{subfigure}
	\begin{subfigure}{0.4\textwidth}
		\includegraphics[width=\textwidth]{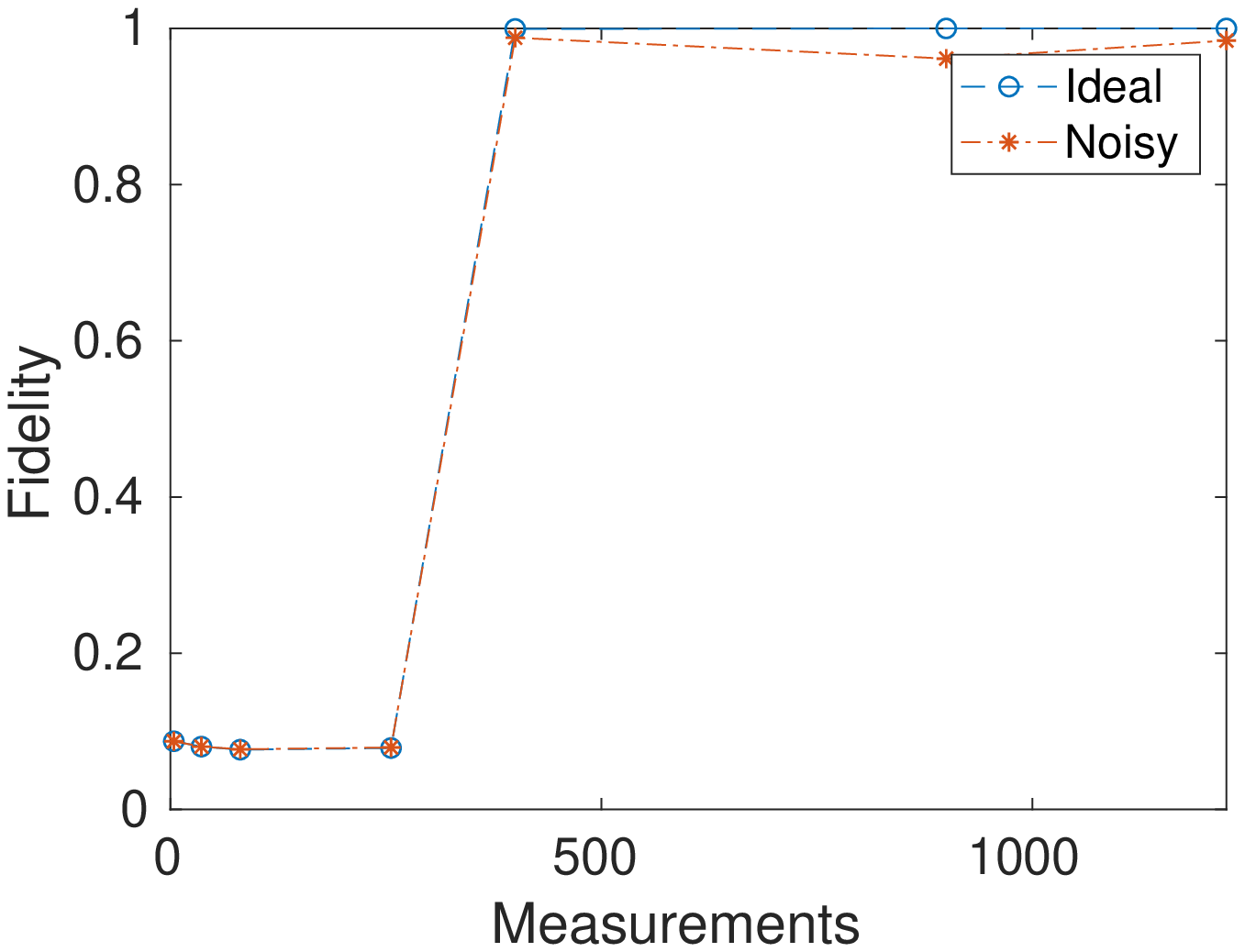}
		\caption{}
		\label{fig:fig4c}
	\end{subfigure}	
	\begin{subfigure}{0.4\textwidth}
		\includegraphics[width=\textwidth]{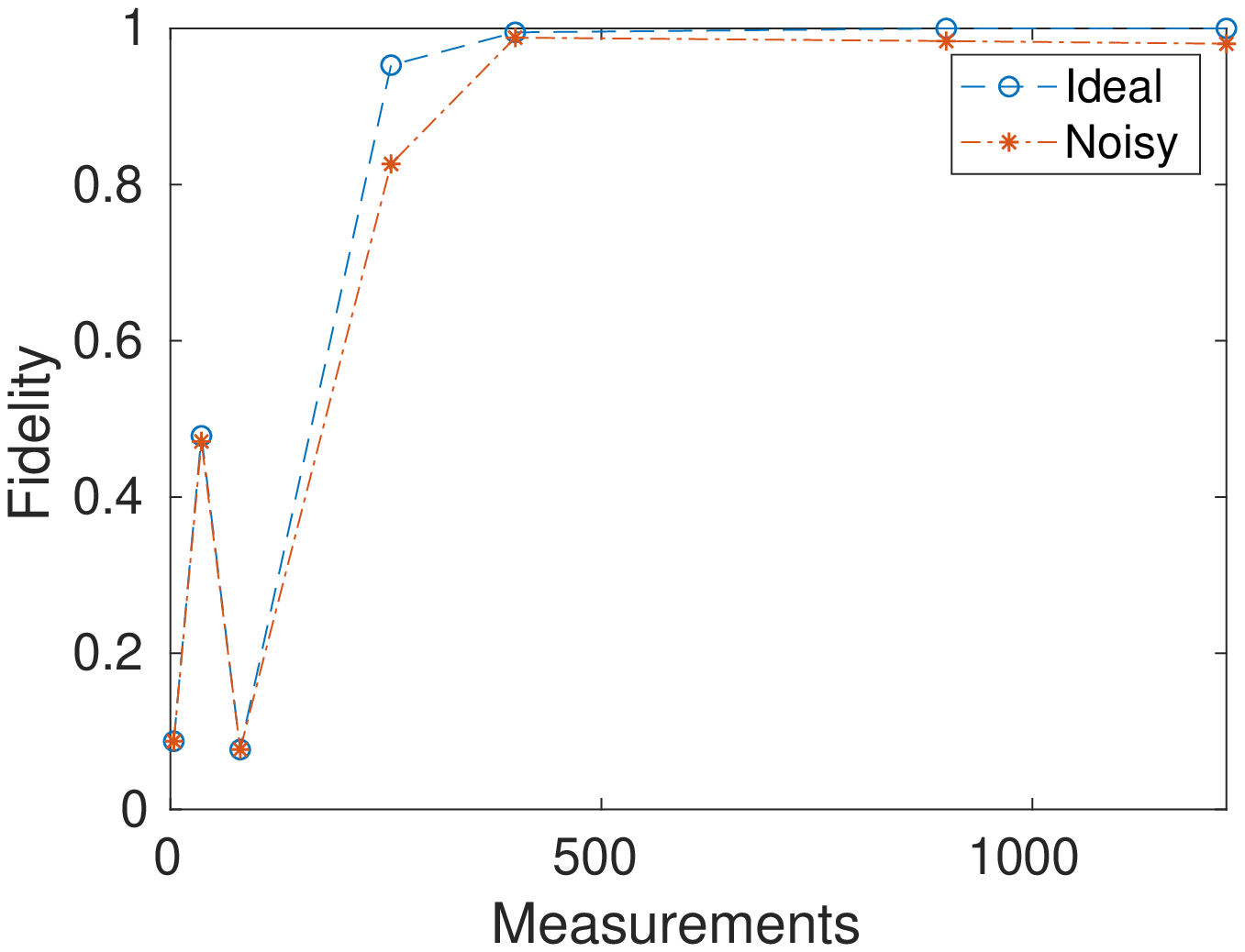}
		\caption{}
		\label{fig:fig4d}
	\end{subfigure}
	\caption{Fidelity to the target state versus number of measured quadratures: (a) Dephased Cat, (b) Hermite-Gauss, (c) NOON and, (d) Squeezed Vacuum.}
	\label{fig:fig4}
\end{figure}
\begin{figure}
	\begin{subfigure}{0.4\textwidth}
		\includegraphics[width=\textwidth]{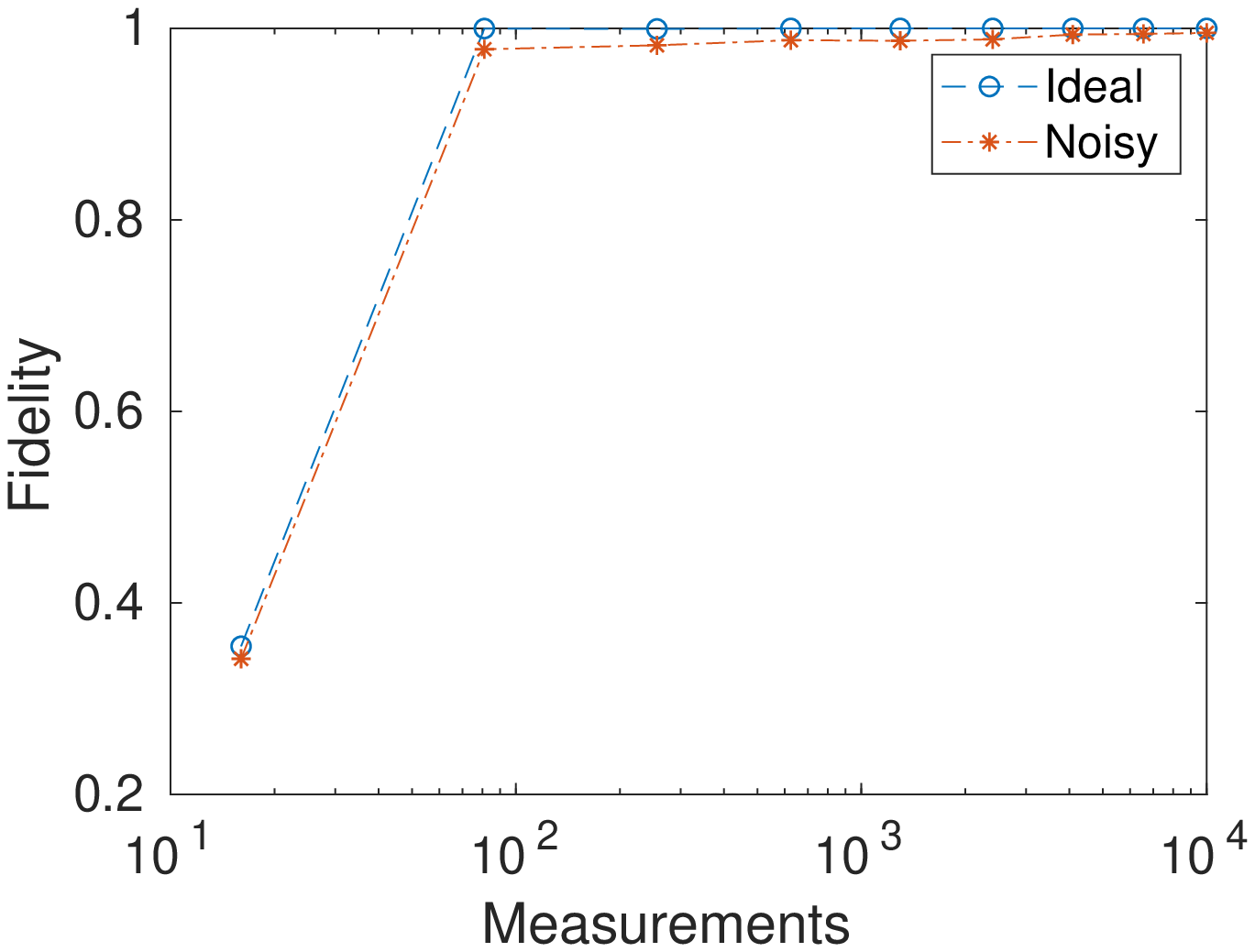}
		\caption{}
		\label{fig:fig5a}
	\end{subfigure}	
	\begin{subfigure}{0.4\textwidth}
		\includegraphics[width=\textwidth]{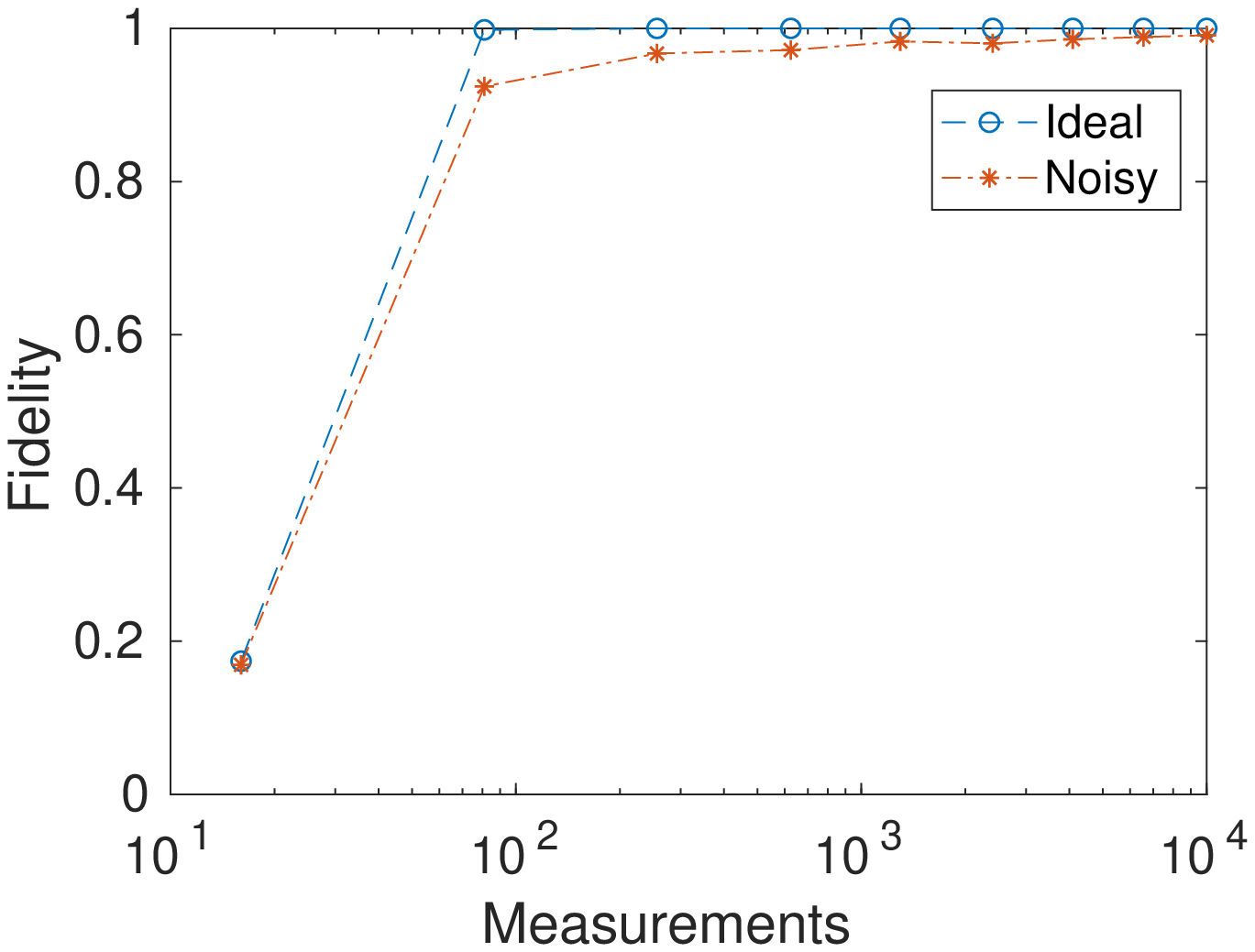}
		\caption{}
		\label{fig:fig5b}
	\end{subfigure}
	\begin{subfigure}{0.4\textwidth}
		\includegraphics[width=\textwidth]{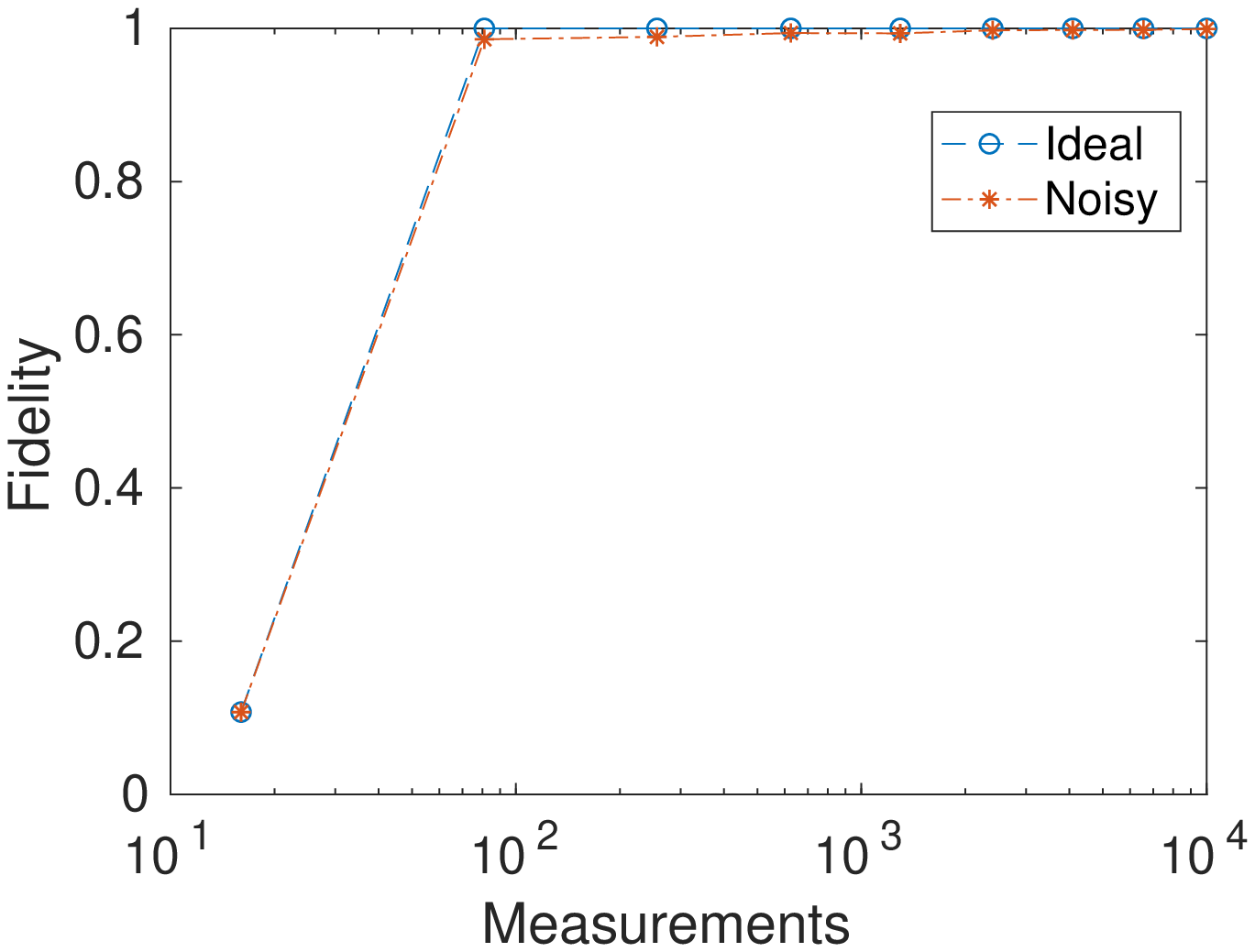}
		\caption{}
		\label{fig:fig5c}
	\end{subfigure}	
	\begin{subfigure}{0.4\textwidth}
		\includegraphics[width=\textwidth]{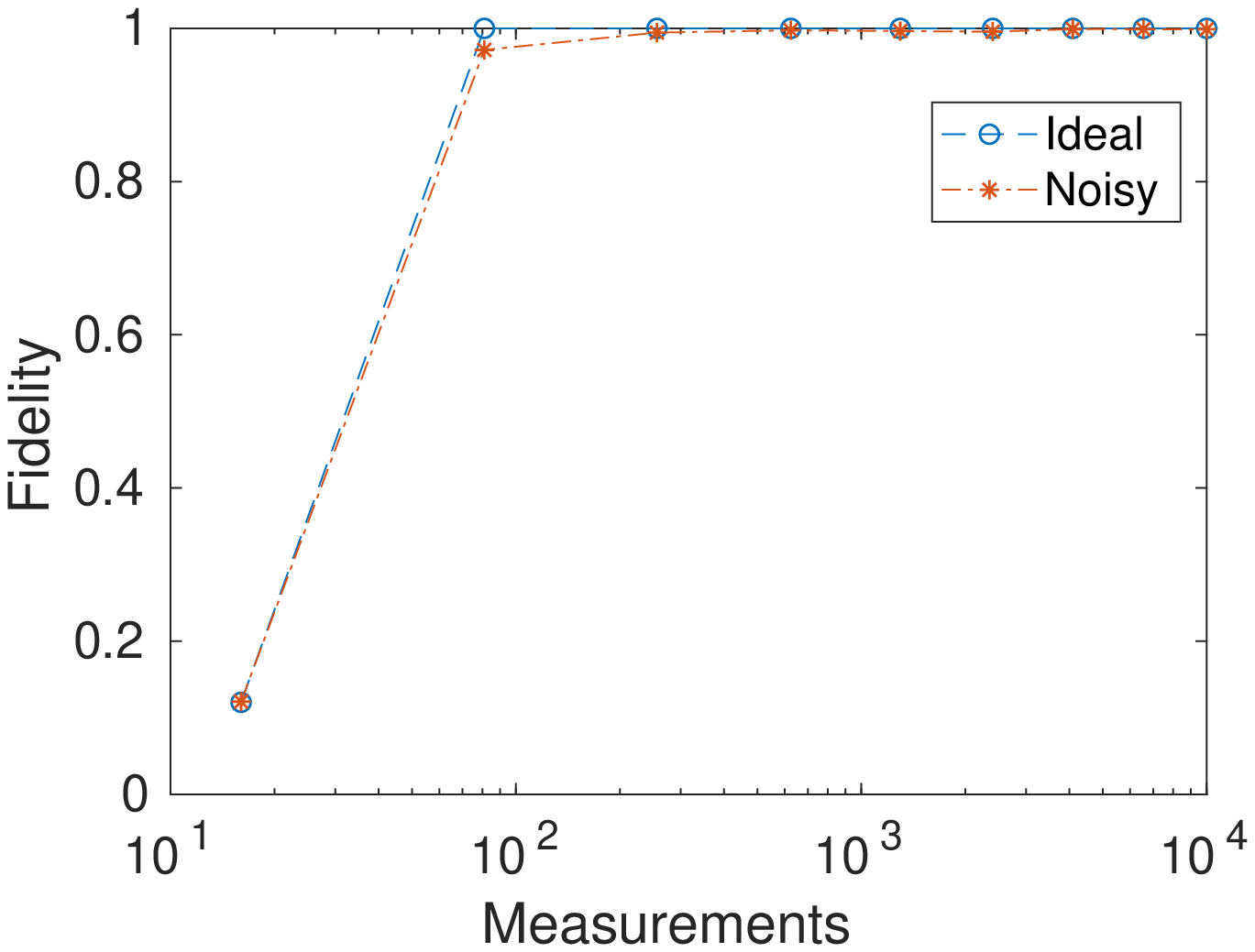}
		\caption{}
		\label{fig:fig4g}
	\end{subfigure}
	\caption{Fidelity to the target state versus number of measured coherent states:  (a) Dephased Cat, (b) Hermite-Gauss, (c) NOON, and (d) Squeezed Vacuum.}
	\label{fig:fig5}
\end{figure}

In Fig.3 we show the reconstruction with quadratures, while in Fig.4 the reconstruction is with coherent states. We simulate noisy measurements by means of a Poissonian
distribution with $10\%$ signal-to-noise ratio. All the reconstructions needed a small number of measurements, even in presence of noise. A high fidelity to the target state is
achieved with about 400 quadratures or 100 coherent states, in all cases. 

To check the unbiasedness of the reconstructed states, we randomly chose ten non-measured observables ($E_i$) and calculated the Shannon entropy of the corresponding
probability vector ($\vec{p}$):
\begin{equation}
S=\sum_{i=1}^{10} -p_i\log_{10}p_i,
\end{equation}
\begin{equation}
p_i=\Tr{\rho E_i}/N,  
\end{equation}
\begin{equation}
N=\sum_{i=1}^{10} p_i.
 \end{equation}
The third line of the SDP program (Eq.\ref{Eq-rov-19}) forces $\vec{p}$ to be the most entropic as possible. As an illustration, we compared the  entropy for the NOON and
Hermite-Gauss states reconstructed, in presence of noise,  with ($S_{unbiased}$) and without ($S_{biased}$) the referred program line.  The results show that our reconstruction
indeed yields a  more entropic state: for the NOON we have $S_{biased}=0.197$ and $S_{unbiased}=0.228$, while for the Hermite-Gauss $S_{biased}=0.333$ and $S_{unbiased}=0.362$.

In the finite Fock basis representation, it is easy to quantify the entanglement of the reconstructed states, by partially transposing the state ($\rho^{T_1}$) and calculating
the Negativity (\cite{Vidal}):
\begin{equation}
\mathcal{N}=\frac{||\rho^{T_1}||_1 - 1}{2}.
\end{equation}
We obtained $\mathcal{N}$(Hermite-Gauss)=0.89,   $\mathcal{N}$(NOON)=0.50,  $\mathcal{N}$(Squeezed Vacuum)=0.25, and  $\mathcal{N}$(Dephased-Cat)=0.24.

\section{Conclusion}

We showed that truncating the Fock basis at number state $n=10$, it is possible to represent continuous variable states with high numerical precision ($10^{-8}$) in the
expectation values. This discrete Fock basis finite representation allowed us to simulate the reconstruction of two-mode states in a small desktop computer. We 
introduced a semidefinite program that performs the tomography of the continuous variable states very efficiently in presence of noise, and yields unbiased states in the
maximum entropy principle sense. The two-mode states we tested, needed the measurement of about 400 quadratures or 100 coherent states to achieve a high
fidelity ($> 0.9$) reconstruction. By taking advantage of the small finite representation, we assessed the entanglement of the states by means of the negativity.

\acknowledgments We acknowledge financial support from the Brazilian agencies FAPEMIG, CAPES, and CNPq INCT-IQ (465469/2014-0).

\end{document}